**Title: Laser patterning of magnonic structure via local crystallization of Yittrium Iron Garnet**

A. Del Giacco, F. Maspero, V. Levati, M. Vitali, E. Albisetti, D. Petti, L. Brambilla, V. Polewczyk, G.Vinai, G. Panaccione, R. Silvani, M. Madami, S. Tacchi, R. Dreyer, S. R. Lake, G. Woltersdorf, G. Schmidt, Riccardo Bertacco*

*Andrea Del Giacco, Federico Maspero, Valerio Levati, Matteo Vitali, Edoardo Albisetti, Daniela Petti, Riccardo Bertacco\**
Dipartimento di Fisica, Politecnico di Milano, Via G. Colombo 81, 20133 Milano (Italia)
* E-mail: riccardo.bertacco@polimi.it

*Luigi Brambilla*
Dipartimento di Chimica, Materiali e Ingegneria Chimica Giulio Natta, Politecnico di Milano, P.za L. da Vinci 32, 20133 Milano (Italia)

*Vincent Polewczyk, Giovanni Vinai, Giancarlo Panaccione*
*Istituto Officina dei Materiali del CNR (CNR-IOM), SS. 14, km 163,5, 34149 – Trieste, Italy*

*Raffaele Silvani, Marco Madami*
*Dipartimento di Fisica e Geologia, Università di Perugia, Via A. Pascoli, 06123 Perugia, Italy*
*Silvia Tacchi*
*Istituto Officina dei Materiali del CNR (CNR-IOM), Unità di Perugia, Via A. Pascoli, 06123 Perugia, Italy*

*Rouven Dreyer, Stephanie R. Lake, Georg Woltersdorf, Georg Schmidt*
*Martin Luther University Halle-Wittenberg, Institute of Physics, Von-Danckelmann-Platz 3, 06120 Halle (Saale), Germany*
*Georg Schmidt also:*
*Martin Luther University Halle-Wittenberg, Interdisziplinäres Zentrum für Materialwissenschaften, Heinrich-Damerow Straße 4, 06120 Halle (Saale), Germany*

**Abstract**

The fabrication and integration of high-quality structures of Yttrium Iron Garnet (YIG) is critical for magnonics. Films with excellent properties are obtained only on single crystal Gadolinium Gallium Garnet (GGG) substrates using high-temperature processes. The subsequent realization of magnonic structures via lithography and etching is not straightforward as it requires a tight control of the edge roughness, to avoid magnon scattering, and planarization in case of multilayer devices.

In this work we describe a different approach based on local laser annealing of amorphous YIG films, avoiding the need for subjecting the entire sample to high thermal budgets and for physical etching. Starting from amorphous and paramagnetic YIG films grown by pulsed laser deposition at room temperature on GGG, a 405 nm laser is used for patterning arbitrary shaped ferrimagnetic structures by local crystallization. In thick films (160 nm) the laser induced surface corrugation prevents the propagation of spin-wave modes in patterned conduits. For thinner films (80 nm) coherent propagation is observed in 1.2 μm wide conduits displaying an attenuation length of 5 μm which is compatible with a damping coefficient of about $5 \cdot 10^{-3}$. Possible routes to achieve damping coefficients compatible with state-of-the art epitaxial YIG films are discussed.

1. Introduction

Magnonics is nowadays considered a promising technology for implementing wave-computing strategies and high-frequency analog signal processing.[1] The much shorter wavelength of spin-waves with respect to free-space propagation of electromagnetic waves paves the way to miniaturization of devices working in the GHz and THz regime.[2][3][4][5][6] Non-linear phenomena can be easily exploited to modulate spin-wave transmission and implement logic, computing or signal processing functionalities.[7][8][9][10][11] Furthermore, spin currents associated to coherent magnon propagation can be used to efficiently drive domain-wall motion, thus opening the way to the local storage of the information carried by spin waves.[12] Nevertheless some bottlenecks still hamper the development of a mature magnonic technology with industrial applications. High insertion losses, the need for bulky electromagnets to bias devices and the difficulties related to the fabrication of devices with long propagation length for SWs are the main obstacles to be overcome.

The low damping coefficient ($\alpha$) of Yttrium Iron Garnet (YIG) is the main reason for the wide usage of this materials in magnonic structures. Record values of $3\cdot10^{-5}$ and $6.5\cdot10^{-5}$ have been reported for bulk single crystals and thin films grown on Gadolinium Gallium Garnet (GGG) substrates.[13][14] Epitaxial films grown by liquid phase epitaxy on GGG, can reach values of $\alpha$ on the order of $1\cdot10^{-4}$. [15][16] Similar values are found in films deposited by Pulsed Laser Deposition on GGG, displaying $\alpha$ as low as $2\cdot10^{-4}$.[17][18][19] Record values of $\alpha = 5.2\cdot10^{-5}$ have been recently reported for 75 nm YIG films grown by sputtering on GGG followed by post-annealing up to 900°C. [20]

Noteworthy, these numbers can be achieved only on GGG(111) single crystal substrates, due to the extremely small lattice mismatch which promotes pseudomorphic growth. So far, attempts to integrate high quality YIG films on silicon using optimized templates failed in reproducing the quality of films grown on GGG single crystals. In the best case, $\alpha$ on the order of $2\cdot10^{-3}$ are reported for YIG films on Si/SiO$_x$.[21] Furthermore, processes for the synthesis of epitaxial YIG films require high temperatures ( > 600°C), either during the growth or post annealing, which are not compatible with the thermal budget acceptable for integration on complementary metal oxides semiconductor (CMOS) platforms.

Another critical aspect to be considered is the fabrication of YIG micro or nanostructures for spin wave propagation. Conventional top-down processes based on lithography and etching must be carefully optimized to avoid YIG degradation and the introduction of extrinsic magnon scattering by edge roughness.[22] In addition, for relatively thick structures, planarization processes are needed to avoid the formation of steps in the metallic antennas at the edges of the conduits or in case of devices requiring additional overlayers. Some solutions to overcome these difficulties have been proposed. Nanochannels for SW propagation have been demonstrated in YIG via dipolar coupling to ferromagnetic metal nanostripes.[23] Recently, a method for the fabrication of free-standing YIG structures has been reported, based on a post-growth annealing inducing the crystallization of three-dimensional amorphous structures defined by room-temperature PLD and lift off using electron beam lithography with only one or two contact points with the GGG substrate.[24] Low values of damping coefficients ($2\cdot10^{-4}$) have been reported, essentially due to the expulsion of dislocation at the bended portions of the suspended structures in proximity to the anchor points on the GGG substrate. The crystallized suspended structures can be released from the GGG substrate and transferred on another substrate such as a silicon wafer, in view of integration with CMOS electronics.[25]

Here we propose a different method for patterning magnonic structures with arbitrary shape embedded in a continuous film of amorphous YIG, by local laser-induced crystallization. Under the right conditions of irradiation, the area exposed to the laser beam undergoes a phase transition from the amorphous (paramagnetic) to the crystalline (ferrimagnetic) phase which allows for spin wave propagation. Arbitrary shaped ferrimagnetic magnonic structures can be patterned by scanning the laser beam according to a pre-defined layout, without need of physical etching of the paramagnetic YIG out of the patterned area and avoiding the exposure of the entire sample (wafer) to a high thermal budget not compatible with integration. Our method is complementary to others starting from epitaxial YIG and producing a local change of the magnetic properties, e.g. via ion-irradiation, so as to introduce a sort of grey-scale magnonics.[26] However, we stress here that in our case we don't need any physical fabrication step of pre-patterning as the local crystallization intrinsically defines the geometry of the magnonic structure embedded in a non-magnetic amorphous film.

The paper is organized as follows. We first describe the laser writing technique by pointing out the critical parameters influencing the performances of patterned structures. Then we report on the optimization of the writing parameters with reference to two YIG thicknesses (160 and 80 nm), showing that only for 80 nm the laser annealing gives rise to corrugation-free patterned areas with uniform magnetic properties leading to a well-defined FMR precession mode. Finally, we demonstrate the suitability of the proposed technique for patterning magnonic structures by investigating spin-wave propagation in rectangular conduits. For conduits patterned in 80 nm thick YIG films we found an attenuation length on the order of 5 microns, compatible with a damping coefficient of about $5 \cdot 10^{-3}$. Possible routes to overcome critical issues currently limiting the propagation length are discussed.

2. Experimental results

2.1 Critical parameters for laser patterning

Laser patterning is performed using the Nanofrazor Explore apparatus by Heidelberg, featuring a 405 nm laser, with minimum spot size $2 \cdot w_0 = 1.2$ μm (where $w_0$ is the waist radius), which is shone onto a substrate mounted on an interferometric stage. A sketch of the process is shown in Figure 1. The intensity (I) at the sample surface can be tuned by varying both the laser power (P) and focus distance (z), corresponding to the sample height with respect to that ensuring the optimal focus (minimum spot size $w_0$). The pattern is created using a raster scanning mode, whose essential parameters are: (i) pixel size $d_{x,y}$ (minimum step of the grid used for the image discretization), (ii) pixel time $\tau_p$ (exposure time for each point in the grid). In this work $d_x=d_y$ so that, from now on, we will just use a unique value for the pixel size: $d_p=d_x=d_y$. The laser is operated in the continuous mode, so that for a given intensity the dose (D) and time-dependent intensity (I(t)) for each pixel size in a line of the pattern can be easily evaluated (see Methods for details) as a function of the writing parameters.

Of course, the local maximum temperature and time profile strongly depend on the optical and thermal properties of film and substrate. In the present case, at 405 nm wavelength the photon energy (3.06 eV) is higher than the gap of YIG and smaller than that of GGG (2.8 eV and 5.6 eV, respectively) so that efficient energy absorption takes place only in the volume of the YIG film.[27][28][29] Since the thickness of films under consideration is smaller (80-160 nm) than the photon penetration depth in YIG at 405 nm (about 0.2 μm) crystallization is expected to start from the GGG substrate, acting as a template for the achievement of single phase structures with (111) orientation. Noteworthy, the maximum temperature reached in the film during the local heating cycle also depends on the film thickness. For film thicknesses lower than the absorption length and transparent substrates, as in our case, the maximum temperature increases with thickness as more heat is absorbed in the film while dissipation channels are unchanged, apart from the lateral heat flow which is not so relevant in this thickness regime. [30]

Finally notice that the shape of the patterned area also influences the laser induced modifications in the film. The ultimate dose in each pixel depends on the superposition of the spots along a single writing line but also on the superpositions of adjacent lines spaced by $d_y$. For the same laser power pixels of a simple one-dimensional line undergo a single heating cycle, while pixels of a true two-dimensional patterned area undergo multiple annealing due to the superposition of adjacent writing lines. With the same laser writing conditions, we thus expect a much stronger impact in circles, squares or wide conduits than in narrow conduits. This must be taken into account when comparing data taken with different experimental techniques on patterned areas adapted to the specific experiment.

2.2. Optimization of patterning conditions

A careful investigation of the impact of the different parameters on the capability of inducing a local crystallization has been carried out by creating 2D arrays of circular and square dots with diameter and side of 20 μm, respectively, spanning different ranges of writing parameters along rows and columns. Their properties have been investigated by optical microscopy, microRaman, electron back scatter diffraction (EBSD), and AFM. In the following we present this analysis for two representative thickness, 160 nm and 80 nm, corresponding to two distinct regimes: (i) a high-thickness regime where the laser irradiation induces a peculiar corrugation with associated spatial modulation of the magnetic properties, (ii) a low-thickness regime where said corrugation disappears and the patterned areas display more uniform magnetic properties.

*2.2.1. 160 nm thick films*

In Figure 1b, we show a typical 2D array created during the optimization of the writing parameters, where we scanned the laser power in the 25-55 mW range on the horizontal axis and $d_p$, between 10 and 90 nm, on the vertical axis. The other parameters were $\tau_p$=70μs, z= 0. In these conditions the line-dose (average intensity) varies from 0.23 to 0.51 mJ/μm$^2$ (13.8 to 30.3 mW/μm$^2$) when going from left to right in the first row of Figure 1b and from 0.23 to 0.026 mJ/μm$^2$ (13.8 to 13.6 mW/μm$^2$) when going from top to bottom in the first column, while the average duration Δt of the heating cycle (constant in each row) goes from 17 to 3.4 ms when moving from the first to the last row. We report this case as it turned out that the power and pixel size are the more effective parameters inducing changes in the local crystal properties. The change in color in the optical microscopy images under white light illumination, from grey to pink and finally light purple, is associated to amorphous-crystalline phase transition, as confirmed by microRaman (see below). Interestingly enough, in the array of Fig. 1b we notice that the induced crystallization is not always uniform in the dot. For some combinations of power and pixel size (e.g. 45mW- 40 nm) we observe that only the upper part of the dots is crystallized, corresponding to the light-purples regions with a tip pointing to the bottom. This is more evident in Fig. 1c, where we report some patterns obtained for P=47.5 mW, $d_p$=30 nm, τp=70 μs, representing a sort of "threshold condition" for crystallization. The phase transition starts at the apex of the tip which corresponds to the initial nucleation center for crystallization (randomly placed in the different dots and probably corresponding to defects increasing local absorption) and then propagates towards the upper part, as the raster scan of the laser proceeds from left to right and from the bottom to the top. Above this threshold, however, the crystallization is more uniform in the dot, as indicated by the quite uniform light-purple color observed in optical microscopy on the right upper part of the array of Figure 2a.

In Figure 1b we observe that local crystallization is favored at high laser power and small pixel size, as these conditions lead to a higher average laser intensity ($I_{av} = D/\Delta t$) during the heating cycle (see Methods). As the crystallization is driven by thermally activated nucleation and growth of crystal grains, the process is mainly determined by the temperature profile during laser heating. The evaluation of the actual temperature evolution in our YIG films requires a careful modelling starting from accurate estimates of the thermal and optical parameters of amorphous or crystalline YIG and GGG, [31] which is beyond the scope of this work. Here we just point out that the maximum temperature is more directly related to the average intensity than to the dose. For this reason, $I_{av}$ can be used as an empirical parameter describing the efficiency of laser induced crystallization for a fixed shape of the patterned area, in agreement with the trend of Figure 1b. This is confirmed also by the fact that we did not find a sizable dependence of crystallization on the pixel time $\tau_p$, as $I_{av}$ does not depends on $\tau_p$ even though the dose depends linearly on it (See Methods).

Figure 1d shows Micro-Raman spectra from dots written in the same conditions of those of the first row of Fig. 1b. As the laser power increases above 40 mW, i.e. in conditions leading to a clear transition towards a light-purple color in the optical images, a YIG $T_{2g}$ feature (C) at about 187.5 cm$^{-1}$ appears, which is a fingerprint of crystalline YIG.[32] The prominent $T_{2g}$ peaks (A,B) at 168.5 and 178.5 cm$^{-1}$ are instead arising from the GGG substrate. The intensity of the YIG $T_{2g}$ peak (C), normalized to that of the adjacent GGG $T_{2g}$ peak (B), is shown in the inset of Figure 1d. These intensities have been calculated by subtracting from each spectrum a reference from an amorphous YIG film and fitting the curve with three peaks to obtain the corresponding amplitudes. We observe a relative increase of the weight of crystalline YIG with respect to the signal from GGG up to 60 mW. For higher laser power the average intensity is so high that crystallization is accompanied by other phenomena, leading to evident morphological transformation (see Figure 2) and chemical modifications (Figure 3).

An intriguing phenomenon observed in 160 nm thick films of YIG is the creation of a peculiar corrugation in areas patterned above threshold, i.e. in conditions leading to crystallization. A regular ripple appears with parallel valleys always perpendicular to the writing directions, as reported in Figure 1e,f,g, where the AFM topography of square areas patterned with a writing direction at 45, 0 and 90 degrees with respect to the [1-10] direction of GGG is shown. This corrugation appears only upon crystallization, as demonstrated in Figure 1e, where we can distinguish the full square area first patterned in the raster-scan mode with P=45 mW ($d_p$=30 nm, $\tau_p$=70 μs, z=0) and then with a second partial raster scan starting from the top-left corner. For this threshold value of writing power, below which we do not observe crystallization and above which crystallization becomes deterministic, the regular ripple appears only in the region patterned twice, where the second localized annealing is capable to induce full crystallization. The effect of the first irradiation is still visible in the AFM topography as a reduction of the films thickness by 5 nm due to the increase of YIG density upon laser annealing, but we have no evidence for crystallization as no magnetic signal was detected in this area (see below). Even though the front of the fully crystallized area is not linear, most probably due to the non-uniform distribution of nucleation centers, the ripple is quite regular, with a peak-to-peak distance of about 1 micron and an amplitude of 10-20 nm. The same kind of corrugation is observed also in figures 1f,g where the writing direction is differently oriented with respect to the GGG crystal axes and we used a power level well above threshold (P=55 mW), while the other parameters are unchanged. Noteworthy, magnetic force microscopy (MFM) (see figure 1, panels h,i,l) indicates that a magnetic contrast is found only in corrugated areas, thus confirming that the ripple is associated to crystallization, as amorphous YIG is paramagnetic. Because the MFM is performed in the lift mode at 200 nm height, we can rule out a strong influence of morphology on the magnetic signal, which must instead be ascribed to the stray field from magnetic domains correlated with the corrugation. This is fully consistent with the magnetic dynamics observed in 160 nm thick films, showing a non-uniform behavior which can be associated with the superposition of magnetic excitations originating from the spatial corrugation (see section 2.3.1). The appearance of this ripple in laser irradiated films has been already reported for other materials and is usually ascribed to the formation of capillary waves due to continuous melting and crystallization of the film during irradiation.[33] The observed corrugation is similar to Laser Induced Periodic Surface Structures which are mainly investigated using fs-lasers but can be also produced with continuous lasers leading to quasi-periodic ripples with low spatial frequency on the order of the laser wavelength.[34][35]

A more detailed analysis of the film transformation induced by laser irradiation at increasing power is reported in Figure 2. In the top row we present optical and AFM images showing the morphology of the patterned areas on YIG films with 160 nm thickness, in the 40-110 mW power range. Below 50 mW the main effect is a laser cleaning of the sample surface which presented some spurious particles. At 47.5 mW (Fig. 2a) we are at threshold so that only a second laser scan can induce the formation of a clear ripple, as reported in Figure 1e,h. From 50 to 60 mW (Fig. 2b) we see the appearance of a long-range corrugation decorated with some crystallites of about 200 nm lateral size and 10 nm height, whose areal density increases with power. At 70 mW (Fig. 2c) we observe a fully developed ripple with perfect coalescence of crystallites and the appearance of elongated protrusions of about 10-15 nm height perpendicular to the laser writing directions, mainly parallel to each other but with some bifurcations, separating regions with very low roughness (< 1 nm rms). These are probably the result of the formation of capillary waves during the solidification of molten material upon irradiation.[33] From 80 to 100 mW (Fig. 2d and 2e corresponding to 80 and 95 mW) we see the appearance of another regime, with holes having increasing depth (50-100 nm) which tend to align in rows. At 110 mW instead (Fig. 2f), we see some elongated bumps with typical height of 200 nm, always presenting some preferential orientation. X-ray absorption spectroscopy (XAS) on irradiated areas of the 160 nm thick film (see Figure 3) reveals that also at 100 mW, the Fe $L_{2,3}$ edge is clearly visible, thus signaling the presence of YIG. However, the line-shape of the $L_{2,3}$ features reported in Figure 3 clearly shows an evolution of the oxidation state of Fe with annealing. At 40 mW and 60 mW the XAS spectrum is identical to that of the as-deposited amorphous film, displaying a ratio between the A and B peaks of the $L_3$ edge which is on the order of 0.5, in nice agreement with the expected value of 0.46 for nominal YIG. [37]. At 80 mW this ratio increases while at 100 mW it becomes larger than one, thus indicating the presence of $Fe^{2+}$, as confirmed by the appearance of the pre-edge feature of $L_2$ (indicated with an arrow in figure 3) in the spectra corresponding to 80 mW and 100 mW. [38] This can be attributed to oxygen vacancies formation at the high temperatures reached in thick YIG films irradiated with 80-110 mW, where we observe a sort of de-wetting of the GGG substrate and YIG accumulation on GGG terraces, associated to a major chemical change of the film which probably involves a surface tension modification.

To summarize this part, writing conditions preserving the YIG stoichiometry and maximizing the crystallinity for 160 nm thick YIG correspond to a pixel size of about 30 nm and a laser power of 60-70 mW, but in this condition the film develops a ripple which results in magnetic inhomogeneity.

*2.2.2. 80 nm thick YIG*

Moving to smaller thickness, namely 80 nm, the formation of the characteristic ripple discussed above is strongly suppressed. The AFM and optical images of amorphous YIG films 80 nm thick, irradiated in the same conditions as for 160 nm thick films, are presented in the bottom row of Figure 2, for a direct comparison. It is immediately clear from the optical microscope images that in this case the irradiated areas with power up to 130 mW never display the same black contrast as the patterns written with 110 mW in 160 nm thick YIG. All transitions are somehow shifted to higher powers. The AFM images shown in panels 2g,h,i, corresponding to 60, 80 and 100 mW of laser power, just show some crystallites with 5-10 nm height on a flat surface and some spurious particles already present on the surface of the film before patterning.

At 110mW we see the appearance of some holes with depth of about 80 nm, compatible with the YIG thickness, which tend to align to the GGG terraces as it happens in the case of thicker films (see panels 2d,e). At 130 mW the roughness definitely increases and we see some YIG bumps with 200 nm height similar to those found for 160 nm thick YIG and reported in panel 2f.

The observed shift at higher laser power of the phase transitions is not surprising as in this thickness range, much smaller than the absorption length, the absorbed power is essentially proportional to the film thickness while thermal dissipation channels are unchanged, so that a lower temperature is expected for the same laser power in 80 nm thick YIG.[30] On the other hand, the formation of the ripple is not simply delayed but mainly suppressed, due to the modification of the peculiar laser absorption, thermodynamic and hydrodynamic conditions leading to this phenomenon in thin films.[33][35]

Also for 80 nm thickness we checked by microRaman that above 70 mW of writing power, the characteristic peak of crystalline YIG at about 187.5 cm$^{-1}$ appears, even though the reduced thickness makes the signal less intense (data not shown). Overall, the laser induced crystallization of 80 nm thick YIG requires a higher laser intensity with respect to thicker films, but the topography of patterns is much more uniform. Furthermore, EBSD (see Figure 5 below) shows that a uniform crystallization can be achieved within a patterned area with 20 µm diameter, thus indicating that thin films are good candidates for an effective excitation and propagation of magnetic perturbations.

2.3 Magnetization dynamics in patterned dots

To assess the magnetization dynamics in patterned dots on 160 and 80 nm thick YIG films we designed a frequency-resolved Magneto Optical Kerr Effect (Super-Nyquist sampling MOKE or SNS-MOKE) experiment with cw RF-excitation provided by a coplanar wave guide (CPW).[39] Different groups of circular dots with 20 µm diameter have been patterned with different writing conditions in the gap region of the CPW (see figure 4a). A RF-current is fed in the CPW to excite magnetization dynamics within the dots via an out-of-plane RF-field in the GHz range and SNS-MOKE is used to sample the amplitude and phase information of the dynamics with its diffraction limited spot size of 300 nm.

2.3.1 Magnetization dynamics of structures patterned in160 nm thick YIG

In Figures 4d,e,f,g,h,i the locally obtained Kerr signals proportional to the real and imaginary part of the dynamic magnetic susceptibility are presented versus the in-plane bias field. In the measurements the magnetic field is aligned perpendicular to the signal line of the CPW while a fixed RF frequency of 2.001 GHz is applied. We investigated patterned dots which correspond to different power levels (50, 60, 70, 80, 90, 100 mW) in an amorphous film with 160 nm thickness as a function of the magnetic bias field applied in plane. The other parameters used for patterning are $\tau_p$=70 µs , $d_p$=30 nm, z=0. Below 60 mW no magnetic signal has been detected, in fair agreement with Raman spectra showing the appearance of a sizable $T_{2g}$ YIG peak only at 60 mW (see Figure 1). Starting from 60 mW of laser power some coherent magnetic signals appear, with increasing intensity up to 80 mW, while above 80 mW the MOKE signal vanishes again. This is consistent with the XAS analysis presented above, showing a substantial modification of the film chemistry starting to develop above 80 mW, with the appearance of a prominent $Fe^{2+}$ signal. However, also from 60 to 80 mW laser power, the line shape of the dynamic MOKE signal is quite complex and does not display a unique resonance, as could be expected for a homogeneous dot excited close to its ferromagnetic resonance conditions (about 22 mT for an excitation frequency of 2 GHz). In fact, the spatially resolved phase images reconstructed from the real and imaginary part data in Figures 4b,c from a dot patterned with 80 mW, taken at 2 GHz with 23 and 27 mT, show a quite inhomogeneous phase distribution which is not compatible with a uniform FMR mode. In both cases the phase of the signal varies in the range of $2\pi$ across the patterned dot. Noteworthy some regions with uniform phase aligned vertically, along the same direction of the ripple described above as the writing direction here is horizontal (see the left part of Figure 4c). Clearly the FMR uniform mode is suppressed by such morphological and magnetic non-uniformity.

This picture is confirmed by micro-focused Brillouin Light Scattering (micro-BLS) investigation of SWs propagation in magnonic conduits patterned in 160 nm thick films by laser irradiation with typical parameters leading to sizable ripple, as discussed below.

2.3.2 Magnetization dynamics of structures patterned in 80 nm thick YIG

In a second set of experiments, we repeated the measurements shown in the previous section for a 80 nm thick YIG layer. The real and imaginary parts of the magnetic susceptibility measured by SNS-MOKE are shown in Figures 5a-f, as a function of in-plane field for different laser power levels between 55 and 80 mW. As in the last set of measurements, an RF-frequency of 2 GHz has been applied to the CPW. The other writing parameters were $\tau_p$=70 µs , $d_p$=30 nm, z=0. A tiny signal with the shape of a single resonance appears at 60 mW, while at 65 mW the clear signature of FMR resonance is visible. Here, the real (dispersive) and imaginary (dissipative) parts of the MOKE signal, proportional to the magnetic susceptibility, show the 90 degrees phase shift expected for FMR. As the writing power becomes larger, the signal starts to deviate from the ideal FMR line shape, while at 80 mW it appears as the superposition of multiple modes. This has a counterpart in the spatially resolved phase images presented in the insets of panels 5c and 5f. While in the case of 65 mW the phase of the MOKE signal is consistent with the FMR uniform mode, for the dot patterned at 80 mW we clearly see that the phase is highly non-uniform, possibly arising from the superposition of higher order modes coupled to some sample inhomogeneity. The analysis of the FMR curves can be used to estimate individual linewidths as well as the saturation magnetization and the Gilbert damping coefficient in a frequency dependent manner. Figure 5g reports the FMR frequency vs the in-plane applied field for the dot patterned with 65 mW, together with a Kittel fit; by assuming $\gamma$ = 28 GHz/T we find that $\mu_0 M_s$ is equal to 0.183±0.001 T, in nice agreement with nominal values for YIG. From the linear fit of the corresponding linewidth of the resonance vs applied field shown in Figure 5h we can extract the damping $\alpha = (5.8\pm0.4)\cdot 10^{-3}$. As we compare this patterned structure with the one irradiated with 75 mW, we obtain just a slightly larger $\alpha = (6.3\pm0.4)\cdot 10^{-3}$, as depicted in Fig. 5i. However, by investigating the zero-frequency linewidth $\mu_0 \Delta H_0$ for these two irradiation powers we see a marked increase, from 34.5±30.7 µT to 174.2±41.2 µT as we enhance the laser power from 65 mW to 75 mW. These values are obtained in the low-frequency regime up to 3.2 GHz since for higher frequencies the interpretation of distinguishable Lorentzian peaks becomes challenging due to the onset of different spin wave modes superimposing the quasi-uniform mode. The enhanced inhomogeneous line broadening for larger laser intensities is in agreement with the observation of the inhomogenous phase distribution in the dot patterned with 75 mW (see Figure 3f) arising from laser induced roughness or even tiny damages in the YIG layer.

Interestingly enough, the dot patterned with 65 mW and displaying an ideal FMR lineshape appears essentially black in the EBSD image of Figure 5j, thus indicating the absence of a well-defined crystal orientation. The fact that it displays a magnetic resonance, however, rules out the possibility that it is just amorphous, as we never measured a SNS-MOKE signal on the unpatterned film, displaying a paramagnetic behavior. We conclude that for 65 mW the film is polycrystalline, with small magnetic crystal grains randomly oriented. Also at 75 mW, when the FMR resonance is well visible, the EBSD image (Figure 5k) shows a polycrystalline composition, with a few small areas having well-defined orientation. A uniform (111) orientation of YIG, compatible with an epitaxial film on the GGG(111) substrate, is found only in the dot patterned with 80 mW, which displays a superposition of modes in MOKE analysis. However, in this case the presence of different spin wave modes superimposing the uniform response prevents the determination of a Gilbert damping parameter $\alpha$ due to indistinguishable resonance lines in the spectrum.

2.4 Spin wave propagation in patterned magnetic conduits

To assess the potential of our technique for fabricating magnonic structures we investigated the propagation of SWs excited by a RF-antenna in conduits patterned by local laser annealing. For 160 nm thick YIG films the patterning creates a characteristic ripple which strongly affects SW propagation (data not shown). SWs display a sort of channeling in regions aligned to the ripple, with a longer propagation distance when the valleys of the ripple are aligned to the wave vector of the excited spin waves. However, in the best case the propagation length is on the order of 1 micron, so that the ripple appears as a severe obstacle to the propagation of SWs. These findings suggests that the corrugation is associated to a magnetic inhomogeneity in agreement with SNS-MOKE measurements of section 2.3.1.

Better results are obtained on 80 nm thick YIG films, as in this case we can achieve full crystallization without sizable corrugation. To avoid the formation of the characteristic ripple perpendicular to the writing direction seen in thicker films, we first patterned elongated conduits with 1.2 μm width and 100 μm length in the 80 nm thick YIG film, using a raster-scan mode with longitudinal writing lines.

Different patterning conditions have been investigated, by varying the focus and the laser power, while keeping $d_P$=30 nm and $t_P$=70 μs fixed. By changing the focus around the optimal condition (z = 0) by ± 2 μm we did not find a sizable change apart from what is expected from the increase of the beam spot size, leading to a decreased intensity and thus to the need to increase the power to retrieve similar results with respect to the optimal focusing conditions. The impact of the laser power is summarized in Figure 6. Panel 6a shows the EBSD map taken on a group of four conduits written with decreasing power, from 100 to 85 mW. The central portion of the conduit displays a single crystal structure with (111) orientation, corresponding to epitaxial YIG on GGG(111), whose width decreases with the laser power until it disappears at 85 mW. Noteworthy, there is a second effect of laser annealing which is visible also at 85 mW. This is evident in Figure 6b, where the EBSD signal (color scale) is superimposed to the secondary electron microscopy (SEM) morphological signal (grey scale). A rectangular conduit, containing the crystallized area and whose width scales with the laser power, clearly emerges from the surroundings, most probably reflecting an amorphous-polycrystalline phase transition. Due to the gaussian profile of the laser beam, only in the central part of the conduit the laser intensity overcomes the threshold for full crystallization while on the sides there are two stripes where the intensity is still enough to promote the creation of polycrystalline grains with random orientation. Of course, the width of these regions scales with the laser power, as the same threshold is achieved for larger distances from the center of the conduit at higher power. This picture is corroborated by the fact that the total width of the conduit seen in SEM images is larger than the beam spot-size, thus indicating that the tails of the laser beam play a role.

In order to investigate SW propagation micro-stripline antennas were fabricated on both sides of the conduits, with groups of 4 conduits crossed by the same stripline connecting the signal and ground of the RF circuit in the ground-signal-ground (GSG) configuration. SW dynamics have been studied by means of micro-BLS measurements applying an external magnetic field of $\mu_0 H_{ext}$= 100 mT parallel to the antenna and to the short axis of the conduits, corresponding to the Damon Eschbach (DE) geometry.

First micro-BLS characterization was performed measuring the SW intensity as a function of excitation frequency in the range between 3 and 12 GHz. A micro-BLS signal is not detected for the conduits written using a power from 40 mW to 70 mW. Interestingly enough, for conduits written with a laser power in the 75-85 mW range, where a crystalline phase with well-defined orientation is not observed in the EBSD, a sizable BLS signal is found. This confirms that also the polycrystalline phase is magnetic and can sustain SW propagation, as already demonstrated by SNS-MOKE measurements. On the other hand, we find that the lowest power for measuring a BLS signal in these conduits (75 mW) is larger than that needed (65 mW) to observe a sizable signal in SNS-MOKE experiments on YIG films with the same thickness (Figure 5). This can be explained by the different shape of the patterned area: the total local dose in a dot with 20 $\mu m$ diameter is larger than that in a conduit with 1.2 $\mu m$ nominal width, corresponding to the laser spot size, due to the multiple superposition of writing lines in the raster-scan mode. We thus expect that the threshold condition for crystallization can be achieved at lower laser power in a large dot with respect to a narrow conduit.

In BLS spectra we observe a single peak whose frequency is about 4.3±0.3 GHz for writing power in the 75-105 mW range and decreases to about 3.6 GHz for a power of 110 and 120 mW (Fig.6h) As it can be seen in Fig. 6f, where the micro-BLS intensity measured at fixed frequency (4.22 GHz) along the transverse x-axis with a step size of 250 nm for the 100 mW sample is reported, this peak corresponds to the fundamental mode of the conduits.

Two dimensional maps of the BLS intensity of the fundamental mode acquired at fixed frequency as a function of the distance from the antenna over an area of about 3×12 $\mu m^2$ with 250 nm step size for the conduits written at 100, 95 and 90 mW are plotted in Figures 6c,d,e, respectively. As it can be seen a micro-BLS signal is detectable up to a distance of about 12 µm. The decay length for the SW intensity, has been estimated from the fit of the micro-BLS intensity profile taken along the y direction at the center of the mode, as shown in figure 6g for the conduit patterned with 100 mW, using the equation $I(y) = I_1 \exp\left(-\frac{2y}{\lambda_D}\right) + I_0$, where $y$ is the position along the waveguide, $I_1$ the initial intensity, and $I_0$ the offset intensity.

As visible in Fig. 6h the largest values of the decay length (approximately 5 µm) are observed for the conduits patterned with a laser power between 90 and 105 mW, while the decay length slightly decreases for lower writing power, consistent with the decreasing width of the single crystal phase in the central portion of the conduits observed in EBSD measurements. Much smaller values of the decay length (down to 1.3 µm) are found for the conduits patterned with 110 and 120 mW. This, together with the decrease of the frequency of the fundamental mode, suggests a major degradation of the magnetic properties at high laser power which can be ascribed both to the chemical modifications observed by XAS and to an increased density of defects and edge corrugation due to grain boundaries. As a matter of fact, EBSD on conduits patterned at 120 mW (data not shown) indicates a larger width of the inner crystal region associated to some interruptions of the conduits, probably arising from local re-amorphization upon over-heating, similarly to what happens in phase-change materials. Notice that the chemical modification seen by XAS in Figure 3 are already evident at 100 mW, where we measured the maximum attenuation length. This apparent inconsistency results from the fact that the laser induced modification does not only depend on the laser power, but also on the shape of the patterned area. For XAS experiments we patterned square pads with 50 µm side, so that the effective heating for the same nominal power was larger than in case of 1.2 µm wide conduits due to the superposition of multiple adjacent lines in the raster scan mode used for writing.

## 3. Discussion

Upon optimization of the writing conditions, good crystalline quality and uniform magnetic properties have been achieved only in thin YIG films (80 nm), as thicker films (160 nm) tend to develop a morphological corrugation which is associated also to a magnetic non-homogeneity. Well-defined FMR lines were found by SNS-MOKE in dots patterned in 80 nm thick YIG films, from which a damping parameter α = (5.8±0.4)·$10^{-3}$ has been estimated, with a low zero-frequency linewidth of (34.5±30) µT, and a saturation magnetization µ$_0$M$_s$=(0.183±0.001)T corresponding to M$_s$ = (146±0.7) kA/m. Spin waves propagation in the DE configuration has been observed by micro-BLS in 1.2 μm wide conduits patterned on 80 nm thick YIG, with an attenuation length of about 5 microns in optimized conditions.

To correlate these experimental findings, we simulated the SWs spectrum for a YIG conduit having a thickness of 80 nm and a width of 1.5 μm (larger than the nominal width of 1.2 μm but corresponding to the width of the crystallized region reported in Fig. 5b for the conduit written with 100 mW) using nominal values for the saturation magnetization and exchange stiffness: $M_s$ =146 kA/m (in agreement with SNS-MOKE) and A=0.42·10$^{-6}$ erg/cm. The simulated frequency of the uniform mode turns out to be 4.4 GHz, in nice agreement with the experimental values (4.3±0.3 GHz for writing power in the 75-105 mW range). This indicates that YIG structures patterned with optimized conditions are characterized by a saturation magnetization and exchange constant within the typical range for high quality YIG films.

The damping parameter within patterned structures can be also extracted from micro-BLS data and compared to that estimated by SNS-MOKE. First, we calculated the SWs group velocity $v_g$ at k=0 from the simulated dispersion curve of the fundamental SW mode in the YIG waveguide, obtaining a value of 683 m/s. Using the relation $\lambda_D = \frac{v_g}{\alpha\omega}$ and the experimental values for the decay length and frequency measured for the sample irradiated at 100mW, we obtain α = 5.6·10$^{-3}$ in good agreement with that found by SNS-MOKE (5.8±0.3 ·10$^{-3}$). This value is above the best values reported for epitaxial films grown on GGG by liquid-phase epitaxy and PLD [13-17] or free-standing structure obtained by re-crystallization [22], both in the low 10$^{-4}$. However, this work represents just a first proof-of-concept indicating the feasibility of local patterning of YIG by laser induced local crystallization, while there is still plenty of room for further optimization.

Here we have shown that, by properly tuning the critical writing parameters suitable conditions for inducing the crystallization without altering the stoichiometry can be found, thus inducing the transition towards a polycrystalline or single-crystal phase, which can sustain magnetic excitations. Noteworthy, these conditions strongly depend on the YIG thickness. In thick films the crystallization is associated to the development of a ripple which is detrimental to the coherent propagation of spin waves but could be engineered to tune the YIG anisotropy and exploit spin-wave channeling. In thinner films shapes with flat topography and uniform magnetic properties sustaining coherent propagation of spin-waves can be patterned. Next steps require a careful investigation of the quality of patterned crystalline structures, which can be obtained in YIG films even thinner than 80 nm and in writing condition avoiding sizable superposition of adjacent writing lines in the currently used raster-scan mode for writing arbitrary shapes. A fine analysis of the edge roughness of patterned areas is another key parameter to avoid extrinsic scattering of spin waves during propagation in conduits. Finally, different combination of doping and substrates should be carefully investigated to tune the micromagnetic properties of patterned structures.

4. Conclusion

To summarize, in this paper we investigated the patterning of magnonic structures via laser induced local crystallization of amorphous YIG films on GGG(111). For thick films (160 nm) the writing conditions leading to the formation of a crystalline and ferrimagnetic phase without sizable modification of the stoichiometry also give rise to a surface ripple with valley and ridges aligned perpendicularly to the writing direction. This in turns causes a magnetic non-homogeneity which is detrimental for uniform magnetic excitations. At reduced thickness this corrugation is suppressed. On 80 nm thick films we patterned single crystal and polycrystalline dots suitable for FMR measurements, from which a damping coefficient α = (5.8±0.4)·$10^{-3}$ was estimated. Magnonic conduits have been patterned on 80 nm films and investigated by micro-BLS, showing an attenuation length on the order of 5 μm for the fundamental DE mode, which is compatible with the damping parameter estimated from FMR. Beyond these promising results, next steps involve a careful optimization of the laser-induced crystallization process in films with controlled thickness below 100 nm, pushing the damping parameter towards the values of epitaxial films on GGG in this thickness range.

5. Methods

*Laser patterning with Nanofrazor Explore:* For local crystallization we exploited the laser-writing option featured by the Nanofrazor Explore tool provided by Heidelberg. The laser is operated in the continuous mode, so that for a given intensity the dose (D) for each point of the grid in the scan of a single line along the x direction can be calculated as follows, assuming a gaussian beam profile: $D = \sum_{-N}^{+N} \frac{2P}{\pi w(z)^2} exp\left(\frac{-2d_x^2 n^2}{w(z)^2}\right) \tau_p$. Here the summation over *n* considers that each point sees an intensity which first increases and then decreases during the motion of the laser beam. N can be safely chosen in such a way that *$Nd_x$= 2w(z)*, while w(z) is the radius of the laser beam where the intensity is reduced by a factor $1/e^2$ with respect to the intensity at the center ($I_0$).

This dose is delivered in a time Δt on the order of $2N\tau_p$, so that the average intensity is $I_{av}$=D/(2N$\tau_p$) while the actual power versus time displays a gaussian profile replicating the spatial profile in the time domain: $I(t) = I(n\tau_p) = \frac{2P}{\pi w(z)^2} exp\left(\frac{-2(d_x/\tau_p)^2 t^2}{w(z)^2}\right)$. As we are dealing with a laser annealing process, this means that each point undergoes a heating cycle of duration $\Delta t = 4w(z)\tau_p/d_x$. Notice that the average intensity increases when decreasing $d_x$, as in the ratio D/Δt the exponential dependence of the dose on $-2(d_x/w(z))^2$ dominates over the increase of Δt which is just proportional to $w(z)/d_x$. These estimates are valid for a single line, while in the patterning of an arbitrary shape by raster scan we must consider also the dose arising from the overlap of the laser spot of the neighbor lines spaced by $d_y$, which in this work is choses equal to $d_x$. This can happen after a considerable time from the heating cycle due the writing of the first line, so that for a two-dimensional area each point undergoes many subsequent thermal cycles. The microstructure of the crystallized phase thus depends also on the shape as discussed in the main text.

*Micro-RAMAN:* The phase transition induced on amorphous YIG films by laser annealing was confirmed with micro-Raman analysis using a Horiba Jobin Yvon LabRam HR800 Raman spectrometer, coupled to an Olympus BX41 microscope. The spectra were recorded in back-scattering geometry, using a 50X objective and a Laser Quantum diode-pumped solid-state laser Torus emitting at 532nm. We set the power of the laser line at 30 mW which allowed us to obtain reliable data without inducing changes or photodegradation of the samples during the measures. Each spectrum has been obtained as an average of three acquisitions lasting 180 seconds each in the range 150-650 cm$^{-1}$.

*X-ray Absorption:* XAS measurements at Fe $L_{2,3}$ edges were taken at the NFFA APE-HE beamline (Elettra synchrotron, Italy) [40] at room temperature in total electron yield (TEY) mode in linear horizontal polarization, with the sample surface at 45° with respect to the incident photon beam. Two orthogonal mechanical slits in front of the endstation were used to minimize the spot dimensions on the patterned samples.

*SNS-MOKE:* In order to map the dynamic properties of the laser written structures within the YIG layer we employ a special type of time-resolved magneto-optical Kerr microscopy (TR-MOKE). In this approach with continuous wave RF-excitation we sample the GHz dynamics with ultrashort laser pulses (with a wavelength of 515 nm) at a repetition rate $f_{rep}$ of 80 MHz locally. The diffraction limited spot size is on the order of 300 nm. In this type of experiment, we use undersampling effects to generate alias frequencies at |$f_{rf}$ – n*$f_{rep}$| whereas the lowest order frequency component is the difference of the rf-frequency with the nearest comb line n of the ultrashort laser pulse train. This frequency within the lowest order Nyquist zone is accessible by a balanced photodetector with MHz bandwidth and can be analyzed by means of Lock-In detection. The direct demodulation at alias frequencies corresponds to an intrinsic frequency modulation and thus no further modulation technique is required. The in-phase and quadrature signals which are demodulated simultaneously at the alias frequency represent the real and imaginary part of the dynamic susceptibility and yield amplitude and phase information. Since we overcome the Nyquist criterion we refer to this technique as Super-Nyquist-Sampling MOKE (SNS-MOKE).[39]

*Brillouin Light Scattering:* Micro-BLS measurements are performed by focusing a single-mode solid-state laser (with a wavelength of 532 nm ) at normal incidence onto the sample using an objective with numerical aperture of 0.75, giving a spatial resolution of about 300 nm. A (3+3)-pass tandem Fabry-Perot interferometer is used to analyze the inelastically scattered light. A nanopositioning stage allows us to position the sample with a precision down to 10 nm on all three axes. A micro-stripline antenna, having a width of 3 μm, is used to excite SWs with a microwave source generator. A dc/ac electrical probe station ranging from dc up to 20 GHz is used for spin-wave pumping. The microwave power is set at +16 dBm on the rf generator output. Micro-BLS measurements were carried out in Damon-Eshbach (DE) geometry applying a magnetic field parallel to the antenna.

Acknowledgements

A. Del Giacco, F. Maspero, R. Bertacco and S. Tacchi acknowledge funds from the EU project MandMEMS, grant 101070536. E. Albisetti acknowledges funding from the European Union's Horizon 2020 research and innovation programme under grant agreement number 948225 (project B3YOND) and from the FARE programme of the Italian Ministry for University and Research (MUR) under grant agreement R20FC3PX8R (project NAMASTE). E. Albisetti and S. Tacchi acknowledge funding from the European Union - Next Generation EU - "PNRR - M4C2, investimento 1.1 - "Fondo PRIN 2022" - TEEPHANY - ThreEE-dimensional Processing tecHnique of mAgNetic crYstals for magnonics and nanomagnetism ID 2022P4485M CUP D53D23001400001". D. Petti acknowledges funding from the European Union - Next Generation EU - "PNRR - M4C2, investimento 1.1 - "Fondo PRIN 2022" - PATH - Patterning of Antiferromagnets for THz operation id 2022ZRLA8F– CUP D53D23002490006" and from Fondazione Cariplo and Fondazione CDP, grant n° 2022-1882. M. Madami, R. Silvani and S.Tacchi acknowledge financial support from the Italian national project TEEPHANY-(PRIN2022P4485M), and NextGenerationEU National Innovation Ecosystem grant ECS00000041−VITALITY (CUP B43C22000470005 and CUP J97G22000170005),

under the Italian Ministry of University and Research (MUR). S. Lake and G. Schmidt acknowledge support by the DFG in the project Harmony

This work has been partially done at Polifab, the micro and nanofabrication facility of Politecnico di Milano.

**Data**

All the data regarding the figure in this work can be find on Zenodo in the repository: "Patterning magnonic structures via laser induced crystallization of Yittrium Iron Garnet_Data " accessible through the following link:

https://doi.org/10.5281/zenodo.10692071

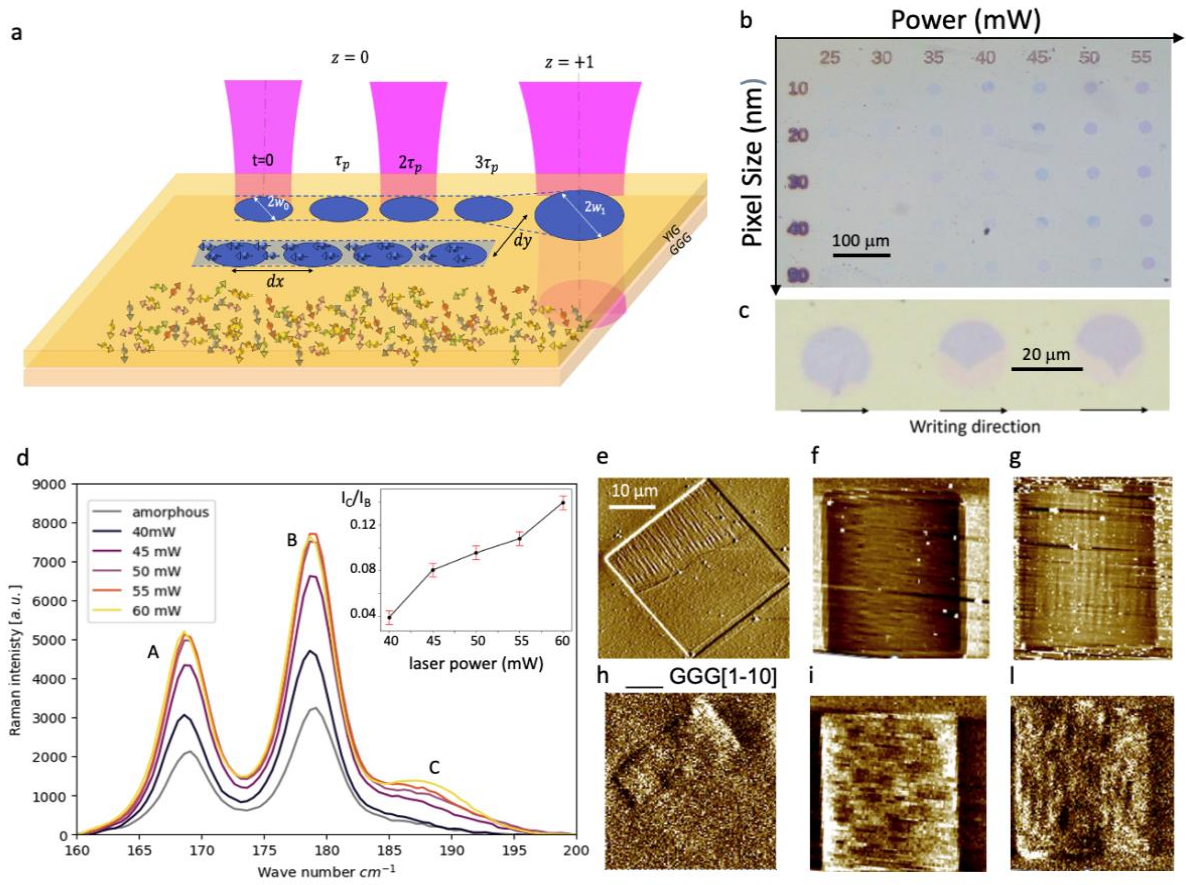

**Figure 1:** a) Sketch of the laser patterning process; b) Optical microscopy of a 2D matrix of 20 microns circular dots patterned with different conditions on 160 nm thick amorphous YIG; c) Zoom on dots patterned at threshold conditions for crystallization. d) microRaman spectra on areas patterned with different laser power; e-g) AFM images and h-l MFM images showing the peculiar ripple perpendicular to the writing direction.

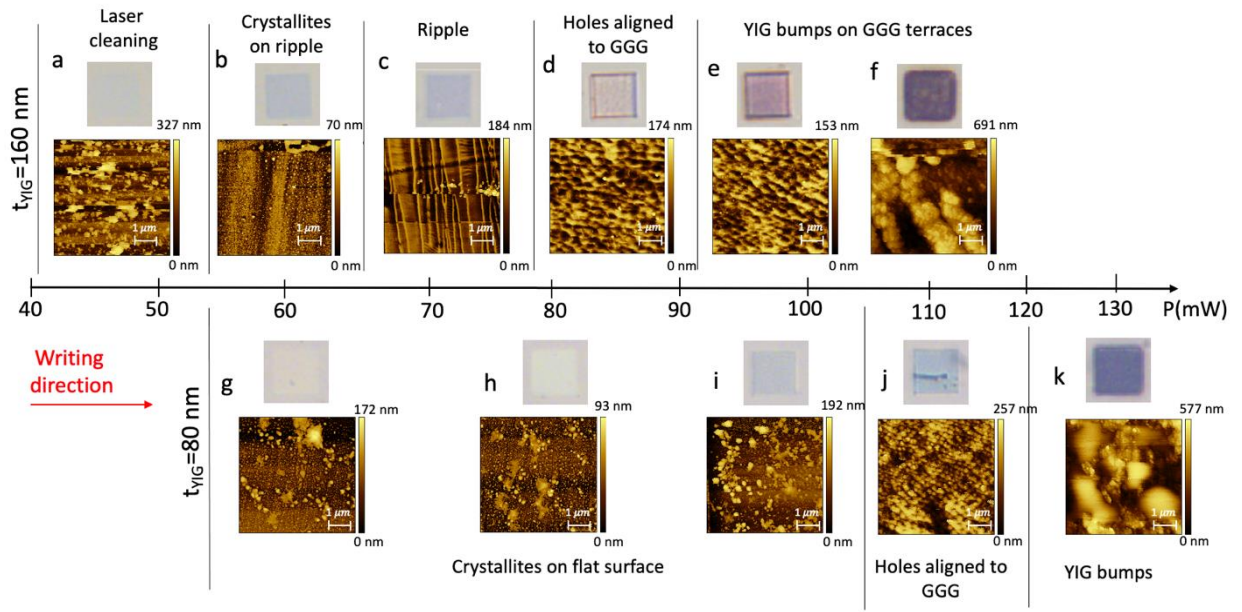

**Figure 2:** Optical (top) and AFM (bottom) images from square patterns with 20 microns side, written on 160 nm (top row) and 80 nm thick (bottom row) amorphous YIG, as a function of the laser power (P) used for writing. The writing direction is horizontal for all patterned areas.

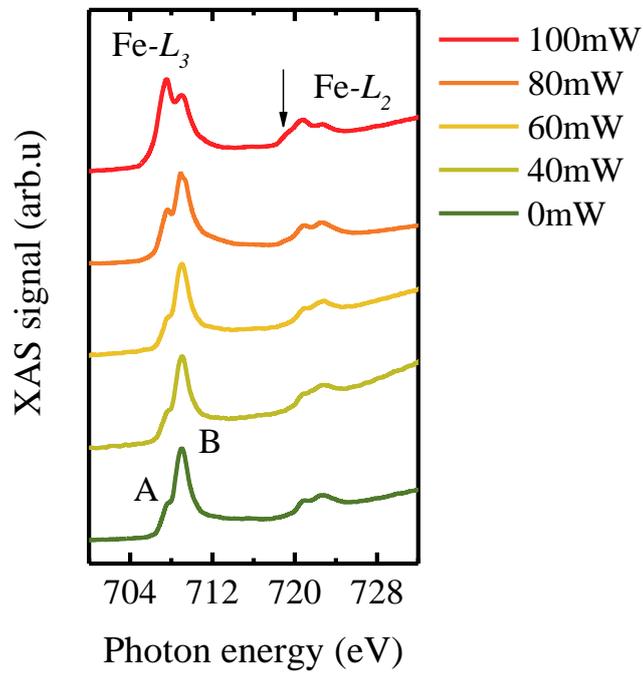

**Figure 3:** XAS spectra from areas patterned on 160 nm thick amorphous YIG at different laser power. Vertical dashed lines correspond to the photon energies of A and B peaks at $L_3$ edge. The vertical arrow indicates the shoulder of the $L_2$ edge which is associated to $Fe^{2+}$.

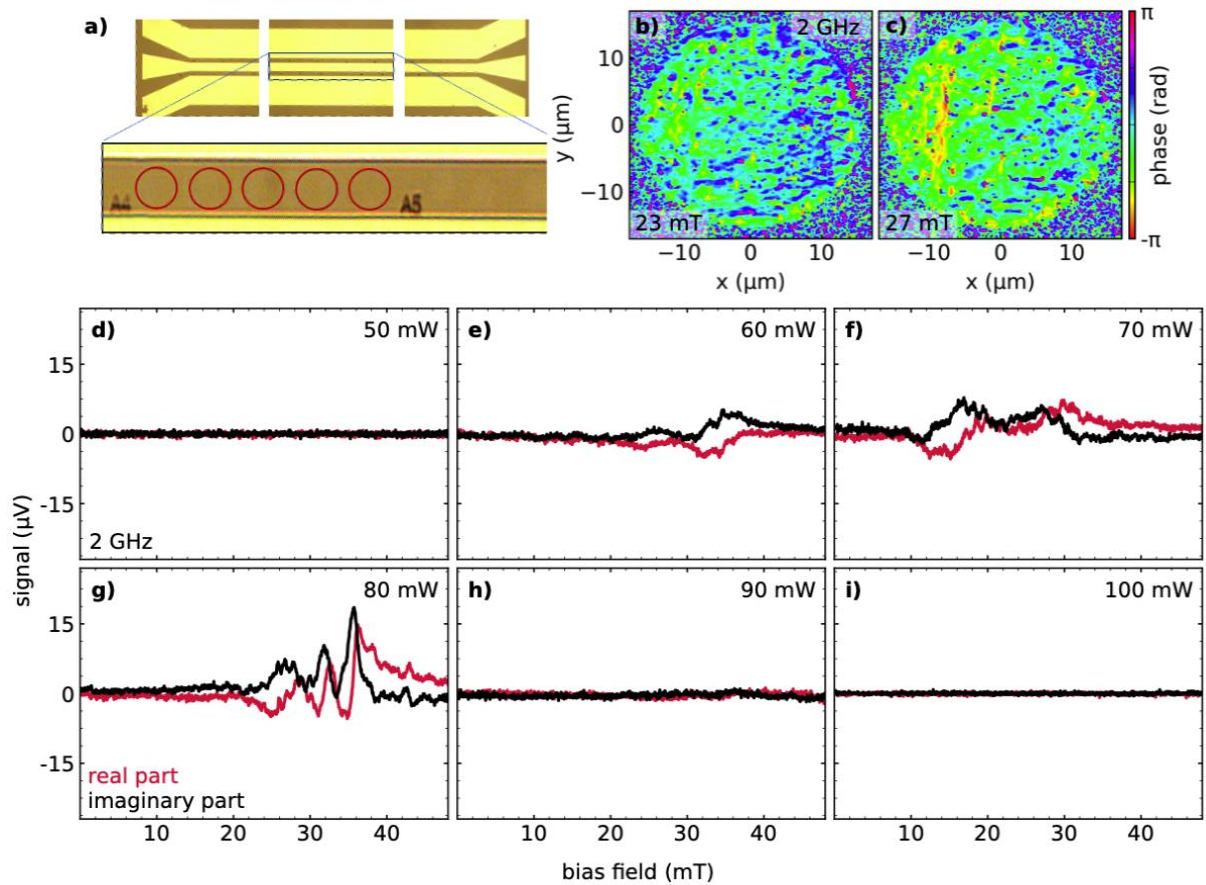

**Figure 4:** TRMOKE analysis on circular dots patterned on 160 nm thick YIG. a) Optical image of the coplanar waveguide (Au/Ti – yellow) fabricated on top of YIG (brown); at the bottom a zoom in the region between signal and ground where series of 5 circular dots with 20 μm diameter (outlined by red circles) have been patterned for each laser power. The bias field was applied along the horizontal x direction. b-c) two-dimensional map of the phase of MOKE signal recorded at 2 GHz on a dot patterned with 80 mW, for 23 and 27 mT of applied field, respectively. d-i) real and imaginary part of the MOKE signal as a function of the applied magnetic field for a fixed excitation frequency of 2 GHz as a function of the writing power reported on the top-right of each panel.

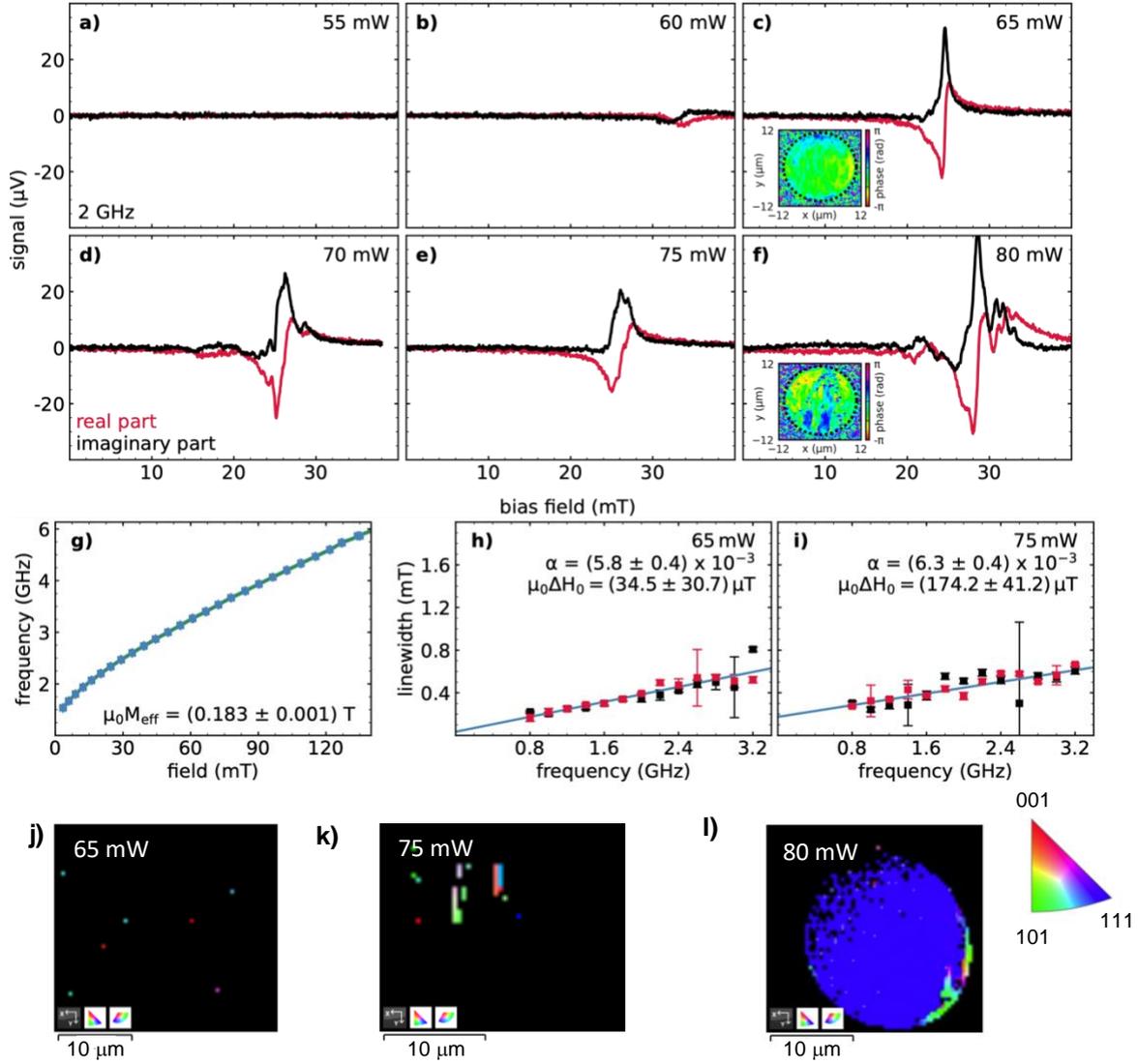

**Figure 5:** TRMOKE data on circular dots patterned at different laser power on 80 nm thick YIG, taken from experiments carried out using the same layout of figure 4a. a-f) real and imaginary part of the TRMOKE signal as a function of the applied magnetic field for a fixed frequency of 2 GHz in the excitation waveguide; in the inset of panels c and f representative maps of the instantaneous phase recorded in the dots are reported. g) resonance frequency vs. applied field of for the dot patterned with 65 mW and showing a clear FMR line-shape; h) plot of the FMR linewidth vs. frequency for the dots patterned with 65 mW; the damping parameter ($\alpha$) has been estimated from the slope of the linear fit (red line). i) same analysis as in panel h for a dot written with 75 mW (panel e). j-k-l) EBS images taken from dots written at 65, 75 and 80 mW.

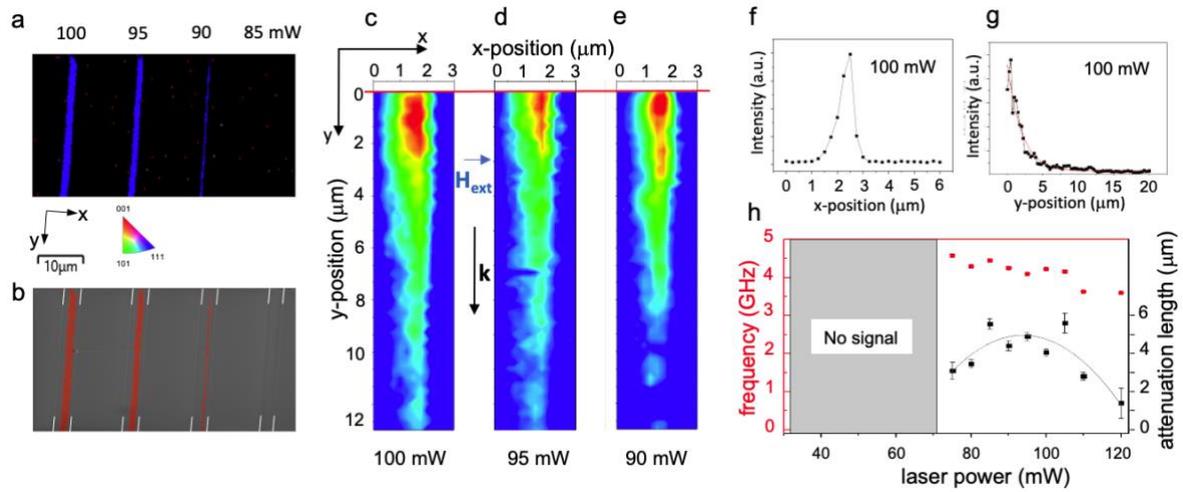

**Figure 6**: BLS analysis of magnonic conduits with nominal width of 1.2 μm patterned on 80 nm thick YIG. a) EBSD image from conduits patterned with a laser power between 85 and 100 mW, the blue color indicates a crystalline zone with (111) orientation; b) EBSD image (red color for (111) orientation), superposed to a conventional SEM image (grey scale) where white segments are used to highlight the borders of the conduits containing the crystallized areas; c-d-e) two-dimensional maps of the BLS intensity on conduits patterned with 100, 95 and 90 mW; on the y axis the distance from the antenna (corresponding to the red line at the top) along the propagation direction (SW wave-vector k) is reported, while x gives the transverse coordinate; f) transverse section of the map in panel c at 6 μm distance from the antenna; g) exponential fit of the BLS intensity decay reported in panel c; h) attenuation length (black dots) and frequency of the propagating mode (red dots) vs. the laser power used for patterning.